\documentclass[onecolumn,secnumarabic, nobibnotes, aps, pra, showpacs]{revtex4-1}
\usepackage{amsmath}
\usepackage{amssymb}
\usepackage{graphicx}

\begin{document}

\title{Quantum fluctuation theorem for dissipative cyclotron motion}
\author{G. S. Agarwal}
\email[Electronic Address:]{girish.agarwal@okstate.edu}
\affiliation{Department of Physics, Oklahoma State
University, Stillwater, Oklahoma 74078 USA}

\author{Sushanta Dattagupta}
\email[Electronic Address:]{sushantad@gmail.com}
\affiliation{Jawaharlal Nehru Centre for Advanced Scientific Research, Bangalore 560 064, India}

\date{\today}

\begin{abstract}
The fluctuation theorems which traditionally have been studied for classical systems have also received substantial attention in the quantum domain over the last decade or so. We report further progress by an explicit analysis of quantum features especially at low temperatures. The example chosen is that of a problem which has proven to be of great interest in the context of Landau diamagnetism viz. the quantized motion of an electron in a magnetic field and in a dissipative environment. It is established from first principles that the quantum work operator W has a Gaussian distribution even though the system under consideration has a four dimensional phase space. The parameter $\alpha$ in the fluctuation theorem $p(W )/p(-W) = exp(\alpha W)$ depends on the system dynamics and has characteristic quantum features, especially at low temperatures. Certain unique low temperature signatures are also discussed.
\end{abstract}

\pacs{05.40.-a, 05.70.Ln, 05.60.Gg}
\maketitle

\affiliation{Department of Physics, Oklahoma State
University, Stillwater, Oklahoma 74078 USA}

\affiliation{Visva-Bharati, Santiniketan 731235, India\\
Honorary Professor, Jawaharlal Nehru Centre for Advanced Scientific Research, Bangalore, India}

\section{Introduction}

The epithet dynamics in the word Thermodynamics is an oxymoron as the subject of Thermodynamics normally deals with macro systems that are in thermal equilibrium at a fixed temperature $T$. Recently however much attention has been focused on open, non-equilibrium systems which are driven by external, time-dependent forces \cite{1gsaprl,2baiesi,3matteo,4mai,5speck1,6ohta,7evans,8hae,9Pro,10bo,11eva,12jar,13jarz,14Talkner,15muk,16dorn,17camp,18esp,19cro,20hor}. A significant such development in the area of non-equilibrium statistical physics is the so-called fluctuation theorems which strive to enunciate relations that mimick similarity with equilibrium systems. In the above-mentioned papers, the fluctuation theorems have been stated in various distinct forms. These were originally done in the context of systems undergoing Brownian motion and interacting with time dependent fields. In an interesting paper Jarzynski and Wojcik had dealt with the statistics of heat exchange between two finite systems initially prepared at different temperatures and had shown that they obey a fluctuation theorem \cite{13jarz}.

Our focus in this paper is on fluctuation theorems for quantum systems, which have also received earlier attention in different forms \cite{14Talkner,15muk,16dorn,17camp,18esp} and reviewed extensively by Campisi, Haenggi and Talkner \cite{17camp} and by Esposito, Harbola and Mukamel \cite{18esp}. Further the major reviews \cite{17camp,18esp} discuss the influence of the heat bath on the system which is assumed to be either weakly or strongly coupled to the bath. However, a crucial difference between the present paper and the earlier mentioned reviews is in the choice of initial conditions. In \cite{17camp} and \cite{18esp}, the combined system and the bath are assumed to be initially in thermal equilibrium described by a canonical density matrix. In our work, based on a derived quantum Langevin equation (from an underlying Hamiltonian in which the system and the bath are strongly coupled), the system is assumed to be in an arbitrary non-equilibrium initial state while the bath is taken to be ever in thermal equilibrium. With the built-in  fluctuation-dissipation relations the system is guaranteed to evolve to a stationary state in which it comes to be in thermal equilibrium with the bath. Besides, for the specific problem of cyclotron motion in the presence of a time-dependent electric field at hand, we are able to provide an exact solution to the Quantum Langevin equation, which is then further analyzed in the context of fluctuation theorems.

In the present paper we would like to develop an approach to a quantum system which in many ways would be parallel to the classical systems \cite{4mai,5speck1,6ohta,7evans,11eva,12jar,21agar3}. As indicated earlier the fluctuation theorems for classical systems are mostly studied using Langevin equations. We follow the same strategy for the quantum case as well i.e we would use the frame work of the Quantum Langevin equations \cite{22newgsabook,23newscu}. Recall that the Boltzmann-Gibbs measure for the probability of an energy configuration $E$ is given by $p(E )\sim \exp(-E/K_B T )$, $T$ being the temperature of the system. The fluctuation theorem that we derived in I for a multivariate Gaussian stochastic process avers that there exists a corresponding statement for the probability measure for the time-dependent work done $W$ due to an external force $f(t)$ applied to a system, otherwise in thermal equilibrium, at a temperature $T $. The statement reads: $p\{W\} / p\{-W\} \sim \exp(\alpha W)$, where $\alpha$  is a positive temperature-dependent factor whose form is by no means universal. In fact, in \cite{21agar3} we showed that only for a one-dimensional process wherein the force acts on a single-component displacement vector does $\alpha$ reduce to $1/(K_B T)$ where $K_B$ is the Boltzmann constant.

Now the question is: can one develop a similar formulation in the quantum case? Following what is known from the field of quantum optics, we adopt quantum Langevin equations to study the fluctuation theorems for quantum systems. In order to bring out the essence of this approach we consider an exactly soluble problem which is of great importance both in the context of the existence of diamagnetism in dissipative quantum mechanics as well as the Hall effect. For this we turn to a well-studied two-dimensional problem of diffusive Brownian motion of a charged particle under the influence of an external magnetic field $B$, applied say along the $z$-direction \cite{24sdgprl}. The two-dimensionality of the cyclotron motion and its relevance to the fluctuation theorems, but for classical systems, has been studied by several authors especially in the over damped limit \cite{25jay}. In addition to being an issue of fundamental interest in `dissipative' Electrodynamics, the problem has application to Condensed Matter phenomena of Diamagnetism and Hall effect (both classical and quantal), as well as to the cyclotron motion in an accelerator. Because the applied magnetic field, through the Lorentz force, does no work on the system, one has to apply a force say in the $x$-direction, to drive the system out of the equilibrium. Further, the problem is essentially two-dimensional in the plane normal to the field. Indeed this configuration is reminiscent of a Hall geometry in which electrons are subjected to transverse magnetic and electric fields.

A few remarks are in order to elaborate on the methodology we adopt. For this it is best to couch the discussion in the Schr\"{o}dinger picture of a density operator instead of the Heisenberg picture of quantum Langevin equation. In our approach the density operator, to start with, is in a factorized form with the system in an arbitrary state while the bath is assumed to be governed by a canonical density matrix appropriate to a constant temperature $T$. The coupling between the system and the bath is then switched on, at time $t = 0$, and one studies the further time-evolution of a `reduced' density operator by `tracing out' (`integrating out' ¨C in the Langevin description) the bath degrees of freedom. The resulting procedure then guarantees that as the time $t$ approaches infinity, the system density operator also acquires the canonical form, governed by the same temperature $T$. This approach is thus completely in accord with the way the canonical ensemble in equilibrium statistical mechanics and the ensuing thermodynamics are worked out.  A further remark is in the context of the externally applied time-dependent field which, in the present context, is a time-dependent electric field. The latter is a classical field representing a measurement process (see the figure below in Sec. II) that couples only with a system (and not a bath) variable -- in the present case, the velocity operator $v(t)$ which, in turn, is modulated by the bath. Note that the aforesaid scenario is exactly the way a spectroscopy experiment is carried out in which the external field plays the role of say, a laser or a synchrotron x-ray beam, and so on. Though the contents of this paragraph are standard text book subject, they are still relevant in highlighting the distinction between our point of view and the one adopted in  the extant reviews on quantum fluctuation theorem \cite{17camp,18esp}.

As mentioned earlier, we examine the work fluctuations by incorporating dissipation into the quantum system when the latter is placed in contact with a quantum bath, in a System-plus-Bath approach, in which the bath degrees of freedom when integrated out, yield frictional force-terms in the equations of motions of the charged particle, thereby explicitly breaking the time-reversal symmetry. While the frictional terms are `systematic', the residual bath interaction shows up through noise terms, yielding a two-dimensional quantum Langevin equation of the following structure \cite{24sdgprl}
\begin{equation}
m\ddot{\vec{q}}+m\gamma \dot{\vec{q}}-\frac{e}{c}(\dot{\vec{q}}\times \vec{B}%
)=\vec{\eta}(t),  \label{Langevin-cyclotron}
\end{equation}
where the first term is the inertial force, the second is the damping term governed by the friction coefficient  $\gamma $, the third is the Lorentz
force-term whereas the right hand side represents the noise-force, which is
a two-dimensional vector. It is interesting to note that the same equation (%
\ref{Langevin-cyclotron}) applies to both classical and quantal domains,
though it is well to remember that while in the classical case, the noise is
usually `white', with delta-correlation in time, it in the quantum case is
necessarily `coloured' subsuming the spectral bath of the environment.
Besides, in quantum mechanics, the phase space variables  are all quantum operators endowed with non-commutation properties.

With Eq. (\ref{Langevin-cyclotron}) in hand we assume that an external force
$\vec{f}\left( t\right) $ is applied along, say the $x$-direction, that
modifies Eq. (\ref{Langevin-cyclotron}) to:
\begin{equation}
m\ddot{\vec{q}}+m\gamma \dot{\vec{q}}-\frac{e}{c}(\dot{\vec{q}}\times \vec{B}%
)=\vec{f}(t)+\vec{\eta}(t).  \label{2}
\end{equation}%
In case the force is due to a time-dependent electric field $\vec{E}\left(
t\right) $,
\begin{equation}
\vec{f}(t)=e \vec{E} \left( t\right).
\end{equation}%
We have earlier shown the validity of the conventional
fluctuation-dissipation theorems for the classical dissipative cyclotron
motion described by Eq. (\ref{2}) \cite{21agar3}. As stated before, here we examine if the fluctuation theorems are applicable to the quantized motion in a dissipative environment. A confining potential which figures in I can be added without any difficulty for the study of Diamagnetism.

The plan of the paper is as follows. In Sec. II we first focus on what is the most appropriate work operator for the quantum description of the cyclotron motion, i.e., we elaborate on our quantum mechanical definition of the work operator. We borrow what is standard in classical electrodynamics and quantum optics and based on this write the work operator. We give the possible experimental scenario which would measure such a work operator. Next, we calculate the work operator from the underlying  quantum Langevin equation. We derive the distribution function for the work operator $W$ and show, quite remarkably that it also follows a Gaussian distribution. The Gaussian distribution ensures the validity of the fluctuation relation $p(W)/p(-W)=\exp(\alpha W)$ in the quantum domain as well but for certain special conditions.  The parameter $\alpha $, unlike the classical case where it is $1/K_B T$ (as in I), now depends on the details of the system and the Planck constant. In Sec. III we present an explicit form of $\alpha $ and find cases when it reduces to the classical result of $1/K_B T$. In Sec. IV we take a specific form of the external drive, endowed with damping and a frequency, in order to provide additional tuning and to delve deeper into unique quantum features. Finally, Sec. V contains a few concluding remarks on future perspectives.

\section{Quantized dissipative cyclotron motion - The probability
distribution of the work operator}

\subsection{DEFINITION OF THE WORK OPERATOR \label{sec21}}

\begin{figure}
  \centering
  \includegraphics[scale=0.5]{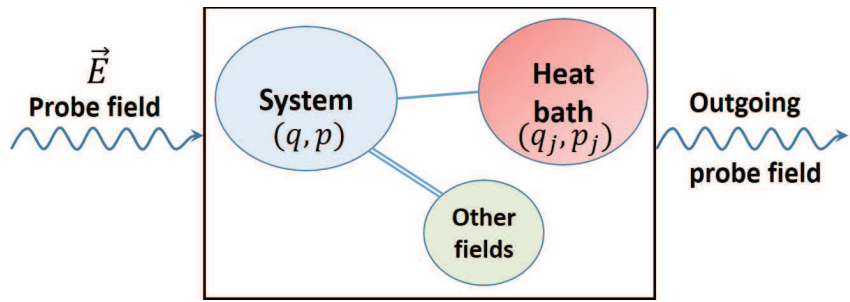}
  \caption{Schematics of the experimental set up.}\label{fig1}
\end{figure}

In order to introduce the work operator we use the approach which has been
extensively used in quantum optics in the interpretation of the experimental
data \cite{26newmol,27newwu}. We sketch it schematically in the Fig. \ref{fig1}.
Other fields in the case would be magnetic field or the confining fields. The probe
field does work on the system which in turn is interacting with a heat bath.
The probe field is an electromagnetic field. The probe field does work on
the charges and thus loses energy. The outgoing probe field and its
fluctuations become related to the work and the fluctuations in work. As
shown in the standard text in electrodynamics \cite{28jackson}, the total rate of
driving work by the probe field $\vec{E}(t)$ on a charged particle is $e\vec{v}(t)\cdot \vec{E}(t)$. Hence the
work operator is taken as
\begin{equation}\label{}
  W(t)=\int_0^{t}e\vec{v}(\tau)\cdot \vec{E}(\tau)d\tau\,,
\end{equation}
where the velocity operator $\vec{v}= \frac{1}{m}(\vec{p}-\frac{e}{c}\vec{A})$.
The $\langle W \rangle$ represents the conversion of electromagnetic energy (from the probe field)
into mechanical work \cite{26newmol,27newwu} which would also depend on the bath characteristics
as the system is in interaction with the bath. Thus the measurement of the rate of change of the energy
of the probe field is a direct measure of the work.

\subsection{QUANTUM LANGEVIN EQUATION\label{secQLE}}

It is well known that adding dissipation in the form of friction terms to a
Lagrangian or a Hamiltonian of a quantum system is fraught with pitfalls as
one may end up with violations of commutation properties or uncertainty
relations between position and momentum operators \cite{22newgsabook,23newscu,29sdgdiffusion}. Hence it
is prudent to begin with a first-principles Lagrangian or Hamiltonian of the
entire `open' system involving the subsystem, the reservoir and their
interaction, and derive an exact quantum Langevin equation. In the context
of quantum fluctuation relations too, we find the same system-plus-bath
approach to be more reasonable rather than juxtaposing an explicitly
time-dependent force from outside onto the by-now established quantum
Langevin equation. As we will see in the sequel the bath coordinate and
momentum variables will appear explicitly in the quantum work operator. This
kind of a first-principles approach to quantum fluctuation relations is new,
to the best of our knowledge.

We therefore start from the Hamiltonian of a charged particle such as an
electron in an external magnetic field and coupled to a harmonic bath,
\textit{a la} Caldeira and Leggett \cite{30leggett} that can be written as
\begin{equation}
H=\frac{1}{2m}\left( \vec{p}-\frac{e}{c}\vec{A}\right) ^{2}+\sum_{j}\left[
\frac{\vec{p}_{j}^{2}}{2m_{j}}+\frac{1}{2}m_{j}\omega _{j}^{2}\left( \vec{q}-%
\frac{C_{j}\vec{q}_{j}}{m_{j}\omega _{j}^{2}}\right) ^{2}\right] \,.
\label{H}
\end{equation}

Here $\vec{p}$'s and $\vec{q}$'s are canonical momenta and coordinates, $%
\vec{A}$ is the vector potential, $m_{j}$ and $\omega _{j}$ are the mass and
the frequency of the $j^{\mathrm{th}}$ bath particle and $C_{j}$ is a
coupling constant. Note the important point that the canonical momentum $%
\vec{p}$ is not the same as the kinematic momentum $m\dot{\vec{q}}$ of Eq. (%
\ref{Langevin-cyclotron}).

From the corresponding Heisenberg equations of motion, and in the presence
of an external force $\vec{f}(t)$, we derive
\begin{equation}
m\dot{\vec{v}}=\frac{e}{c}(\vec{v}\times \vec{B})-m\int_{0}^{t}dt^{\prime
}\gamma (t-t^{\prime })\vec{v}(t^{\prime })+\vec{\eta}(t)+\vec{f}(t),
\label{langevin-cyclo-sd-new}
\end{equation}%
where
\begin{equation}
\gamma (t)=\sum_{j}\frac{C_{j}^{2}}{m_{j}\omega _{j}^{2}}\cos \omega _{j}t\,,
\end{equation}%
and
\begin{equation}
\vec{\eta}(t)=\sum_{j}\bigg\{C_{j}\left[ \vec{q}_{j}(0)-\frac{C_{j}\vec{q}(0)%
}{m_{j}\omega _{j}^{2}}\right] \cos \omega _{j}t+\frac{C_{j}\vec{p}_{j}(0)}{%
m_{j}\omega _{j}}\sin \omega _{j}t\bigg\}\,.
\end{equation}%
The external drive $\vec{f}(t)$ will be reintroduced later, in Eq. (\ref{19}%
) below, but is now dropped in the sequel for the sake of simplicity.


In component form Eq. (\ref{langevin-cyclo-sd-new}) can be written as
\begin{eqnarray}
\dot{v}_{x} &=&\omega _{c}v_{y}-\int_{0}^{t}dt^{\prime }\gamma (t-t^{\prime
})v_{x}(t^{\prime })+\eta _{x}(t), \label{9}\\
\dot{v}_{y} &=&\omega _{c}v_{x}-\int_{0}^{t}dt^{\prime }\gamma (t-t^{\prime
})v_{y}(t^{\prime })+\eta _{y}(t)\,. \label{10}
\end{eqnarray}%
Introducing $v_{\pm}=v_{x}(t)\pm iv_{y}(t)$, and $\eta _{+}(t)=\eta _{x}+i\eta
_{y}$, we have
\begin{equation}
\dot{v}_{+}(t)=i\omega _{c}v_{+}(t)-\int_{0}^{t}dt^{\prime }\gamma
(t-t^{\prime })v_{+}(t^{\prime })+\eta _{+}(t)\,,
\end{equation}%
which, upon Laplace transforming, yields
\begin{equation}
(z+i\omega _{c}+\tilde{\gamma}(z))\tilde{v}_{+}(z)=v_{+}(0)+\tilde{\eta}%
_{+}(z)/m\,,  \label{lap}
\end{equation}%
and hence,
\begin{equation}
\tilde{v}_{+}(z)=A(z)[v_{+}(0)+\tilde{\eta}_{+}(z)/m],
\end{equation}%
where
\begin{equation}
A(z)=(z+i\omega _{c}+\tilde{\gamma}(z))^{-1}\,.  \label{13}
\end{equation}%
Formally,
\begin{equation}
v_{+}(t)=A(t)v_{+}(0)+\sum_{j}\bigg\{C_{j}\left( \xi _{j}(0)-\frac{C_{j}\xi
(0)}{m_{j}\omega _{j}^{2}}\right) B_{j}(t)+C_{j}\frac{p_{+j}(0)}{m_{j}\omega
_{j}}D(t)\bigg\}\,,  \label{formally}
\end{equation}%
where
\begin{eqnarray}
A(t) &=&\mathcal{L}^{-1}\left\{ A(z)\right\} \,,  \label{A} \\
B(t) &=&\mathcal{L}^{-1}\left\{ \frac{A(z)z}{z^{2}+\omega _{j}^{2}}\right\}
\,,  \label{bt} \\
D(t) &=&\mathcal{L}^{-1}\left\{ \frac{A(z)\omega _{j}}{z^{2}+\omega _{j}^{2}}%
\right\} \,,  \label{dt}
\end{eqnarray}%
where $\mathcal{L}^{-1}\left\{ {}\right\} $ denotes inverse Laplace
transforms. Also
\begin{equation}
\xi _{j}(0)=x_{j}(0)+iy_{j}(0)\,,~~\xi (0)=x(0)+iy(0)\,.
\end{equation}%
In the presence of an external force $\vec{f}(t)=(f_{x}(t),f_{y}(t))$,
\begin{equation}
v_{+}(t)=A(t)v_{+}(0)+\sum_{j}\bigg\{C_{j}\left( \xi _{j}(0)-\frac{C_{j}\xi
(0)}{m_{j}\omega _{j}^{2}}\right) B_{j}(t)+C_{j}\frac{p_{+j}(0)}{m_{j}\omega
_{j}}D_{j}(t)\bigg\}+F_{+}(t),  \label{19}
\end{equation}%
where
\begin{eqnarray}
p_{\pm j}(0) &=&p_{xj}(0)\pm ip_{yj}(0)\,, \\
F_{+}(t) &=&\mathcal{L}^{-1}\left\{ A(z)f_{+}(z)/m\right\} \,,  \label{Ft}
\end{eqnarray}%
with
\begin{equation}
f_{+}(z)=\int_{0}^{\infty }dte^{-zt}(f_{x}(t)+if_{y}(t))\,.
\end{equation}%
The `work operator' $W(t)$ is given by
\begin{equation}
W(t)=\int_{0}^{t}dt_{1}(f_{x}(t_{1})v_{x}(t_{1})+f_{y}(t_{1})v_{y}(t_{1}))\,.
\end{equation}%
Because the work operator $W(t)$ contains the velocity operator (under a
time-integral) that does not commute with itself at different times, the
complete time-dependent solution for the velocity operator is required for
the evaluation of the fluctuation in the work operator.

Assuming for simplicity that the force is applied along the $x$-direction $%
(f_{x}(t)=f(t))$,
\begin{eqnarray}
W(t) &=&\int_{0}^{t}dt_{1}f(t_{1})v_{x}(t_{1})=\frac{1}{2}[\alpha
(t)v_{+}(0)+\bar{\alpha}(t)v_{-}(0)]+\phi (t)  \notag \\
&+&\frac{1}{2}\sum_{j}\bigg\{C_{j}\left( \xi _{j}(0)-\frac{C_{j}\xi (0)}{%
m_{j}\omega _{j}^{2}}\right) \beta _{j}(t)+\frac{p_{+j}(0)}{m_{j}\omega _{j}}%
\delta _{j}(t)+C_{j}\left( \xi _{j}^{\ast }(0)-\frac{C_{j}\xi ^{\ast }(0)}{%
m_{j}\omega _{j}^{2}}\right) \bar{\beta}_{j}(t)+\frac{p_{-j}(0)}{m_{j}\omega
_{j}}\bar{\delta}_{j}(t)\bigg\}\,,  \label{Wt}
\end{eqnarray}%
where
\begin{eqnarray}
\phi (t) &=&\int_{0}^{t}dt_{1}f(t_{1})F(t)\,,  \label{25} \\
F(t) &=&\mathcal{L}^{-1}\left\{ A(z)f(z)/m\right\}\,,    \label{26} \\
\alpha (t) &=&\int_{0}^{t}dt_{1}f(t_{1})A(t_{1})\,,  \\
\bar{\alpha}(t) &=&\int_{0}^{t}dt_{1}f(t_{1})\bar{A}(t_{1})\,,  \\
\beta _{j}(t) &=&\int_{0}^{t}dt_{1}f(t_{1})B_{j}(t_{1})\,,  \\
\bar{\beta}_{j}(t) &=&\int_{0}^{t}dt_{1}f(t_{1})\bar{B}_{j}(t_{1})\,,  \\
\delta _{j}(t) &=&\int_{0}^{t}dt_{1}f(t_{1})D_{j}(t_{1})\,,  \\
\bar{\delta}_{j}(t) &=&\int_{0}^{t}dt_{1}f(t_{1})\bar{D}_{j}(t_{1})\,.
\end{eqnarray}%
The barred quantities are obtained by replacing $A(z)$ in Eq. (\ref{13}) by $%
\bar{A}(z)$, i.e. by replacing $\omega _{c}$ by $-\omega _{c}$. Note also that $\alpha(t)$
is distinct from the parameter $\alpha$ in the FD theorem. Thus, the
work operator $W(t)$ explicitly depends on the dynamical variables of the
heat bath, albeit through their initial values, justifying a
first-principles formulation, as mentioned at the outset.
Note further that the work operator (\ref{Wt}) has three distinct contributions (a)
from the initial quantum distribution of the charge particle's velocities (b) the
forcing term $\Phi(t)$ (c) bath terms. For an isolated system the bath terms drop out. The fluctuation
theorems would then depend on the initial distribution.

\subsection{PROBABILITY DISTRIBUTION OF WORK \label{sec23}}

In order to obtain the distribution of $W(t)$, we define the characteristic
function as
\begin{equation}
C_{W}(t)=\langle e^{-ihW(t)}\rangle \,,  \label{char}
\end{equation}%
So that its Fourier transform w.r.t. $h$ (not to be confused with the Planck constant)
would give the probability distribution.
We have
\begin{eqnarray}
C_{W}\left( t\right)  &=&\bigg\langle\exp \bigg\{\frac{ih}{2}\sum_{j}\bigg\{%
C_{j}\left( \xi _{j}(0)-\frac{C_{j}\xi (0)}{m_{j}\omega _{j}^{2}}\right)
\beta _{j}(t)+C_{j}\frac{p_{+j}(0)}{m_{j}\omega _{j}}\delta _{j}(t)  \notag
\\
&+&C_{j}\left( \xi _{j}^{\ast }(0)-\frac{C_{j}\xi ^{\ast }(0)}{m_{j}\omega
_{j}^{2}}\right) \bar{\beta}_{j}(t)+C_{j}\frac{p_{-j}(0)}{m_{j}\omega _{j}}%
\bar{\delta}(t)\bigg\}\bigg\}\bigg\rangle C_{W}^{(0)}(t) e^{ih\phi
(t)}\,,  \label{34}
\end{eqnarray}%
where
\begin{equation}\label{}
C_{W}^{(0)}(t)=\bigg\langle \exp\bigg\{-\frac{1}{2} i h [\alpha(t)v_{+}(0)+\bar{\alpha}(t)v_{-}(0)]\bigg\}\bigg\rangle \,.
\end{equation}
Note that $C_{W}^{(0)}(t)$ is the contribution to the characteristic function coming solely from the initial state of the charged particle.
In what follows we concentrate only on the heat bath contributions. We would thus drop the term $C_{W}^{(0)}$ from
further considerations. It should be borne in mind that the final results for the work and work fluctuations would have
contributions both from the bath part of Eq. (\ref{34}) and the initial value distribution. After dropping $C_{W}^{(0)}$ and
averaging over the bath contributions we get
\begin{equation}
C_{W}(t)=\exp \bigg\{-\frac{h^{2}}{2}\sum_{j}\frac{\hbar C_{j}^{2}}{m_{j}\omega
_{j}}\left( \langle n(\omega _{j})\rangle +\frac{1}{2}\right) [\beta
_{j}^{2}(t)+\bar{\beta}_{j}^{2}(t)+\delta _{j}^{2}(t)+\bar{\delta}%
_{j}^{2}(t)]\bigg\}e^{ih\phi (t)}\,,  \label{evident}
\end{equation}%
where $\langle n(\omega _{j})\rangle $ is the Bose-Einstein factor. We first note that the external 
field contribution $e^{ih\phi (t)}$ factorizes. This is due to the linearity of the underlying Eqs. (\ref{9}) and (\ref{10}),
althrough $\phi (t)$ depends on the bath coupling via $A(z)$ [see Eqs. (\ref{13}) and (\ref{Ft})].
The Gaussianness of $C_{W}\left( t\right) $ emerges from the fact that the bath
operators are taken to be bosonic thereby rendering the bath as harmonic in
addition to assuming the coupling between the system and the bath to be
linear. Though nonlinearity in the system variables has been considered by
earlier authors, it has been extensively argued by Caldeira and Leggett that
the assumption a harmonic bath with a concomitant linear coupling as far as
the bath coordinates are concerned does not vitiate the generality of the
Langevin framework of a quantum dissipative system \cite{30leggett}.

From the characteristic function in Eq. (\ref{evident}) the probability
distribution of the work can be explicitly calculated as%
\begin{equation}
p\left( W\right) =\frac{1}{\sqrt{2\pi }\sigma }\exp \left\{ -\frac{\left(
W-\left\langle W\right\rangle \right) ^{2}}{2\sigma ^{2}}\right\} ,
\end{equation}%
where the variance is the fluctuation in the work operator:%
\begin{equation}
\sigma ^{2}=\left\langle \left( W-\left\langle W\right\rangle \right)
^{2}\right\rangle \,.
\end{equation}%
The expression in Eq. (\ref{evident}) is still a complicated one for
practical use and needs further simplification, as carried out below. However the parameter
$\alpha$ in the fluctuation theorem is $\alpha =2\langle W \rangle/\sigma^2$. We add that in our
`Non-equilibrium' approach we deal with fluctuations but these fluctuations are not static but time-dependent 
and inclusive of dissipative effects. Thus work fluctuations are related to the fluctuations of velocity operator
in the time-domain that depend on dissipative parameters such as friction. It is true that thermodynamic free energy
is also connected with fluctuations - recall that energy fluctuation in equilibrium statistical mechanics is 
related to the Heat Capacity which in turn, can be derived from the second (temperature-) derivative of the Helmholtz Free Energy $F$,
but these energy fluctuations have neither time-nor frictional connotation. This point of view is in conformity with the very notion of 
Thermodynamics in which the system is supposed to be in canonical equilibrium with instantaneous $H(t)$ at each point in time. Only 
in the steady state our $dW(=W-\langle W \rangle)$ can be equated with $-dF$, for an isothermal thermodynamic process.

\section{Explicit results for ohmic dissipation}

In this section we employ the often-used form for the bath spectral function
that goes under the name of ohmic dissipation. We remind the reader that it
is for the ohmic dissipation that the quantum Langevin equation is
describable in terms of a constant friction though the process remains
non-Markovian \cite{29sdgdiffusion}.

Our first task is to calculate the mean work done which, from Eq. (\ref{char}%
), can be expressed as
\begin{equation}
\langle W(t)\rangle =\lim_{h\rightarrow 0}\frac{1}{i}\frac{d}{dh}C_{W}(t)\,.
\end{equation}%
Evidently, from Eq. (\ref{evident}),
\begin{equation}
\langle W(t)\rangle =\phi (t)\,,
\end{equation}%
which, from Eqs. (\ref{25}) and (\ref{26}), yields
\begin{equation}
\langle W(t)\rangle =\frac{1}{m}\int_{0}^{t}dt_{1}%
\int_{0}^{t_{1}}dt_{2}f(t_{1})f(t_{2})e^{-\gamma (t_{1}-t_{2})}\cos \omega
_{c}(t_{1}-t_{2})\,,  \label{40}
\end{equation}%
where we have employed the fact that for Ohmic dissipation the friction is a
constant $\gamma $ and hence
\begin{equation}
A(z)=(z+i\omega _{c}+\gamma )^{-1}\,.
\end{equation}%
Thus the mean work operator has the same form as in the classical case (see
I) and is devoid of the Planck constant!

We now turn to the evaluation of the mean-squared work operator $\langle
W^{2}(t)\rangle $. From Eq. (\ref{char}) it is evident that
\begin{equation}
\langle W^{2}(t)\rangle =-\lim_{h\rightarrow 0}\frac{d^{2}}{dh^{2}}%
C_{W}(t)\,.
\end{equation}%
From Eq. (\ref{evident}),
\begin{equation}
\langle W^{2}(t)\rangle =\phi ^{2}(t)+\sum_{j}\frac{\hbar C_{j}^{2}}{%
m_{j}\omega _{j}}\left( \langle n(\omega _{j})\rangle +\frac{1}{2}\right)
\times [|\beta _{j}^{2}(t)|+|\bar{\beta}_{j}^{2}(t)|+|\delta _{j}^{2}(t)|+|%
\bar{\delta}_{j}^{2}(t)|]\,.
\end{equation}%
Because $\phi ^{2}(t)$ is of quartic order in the external force it may be
dropped from further considerations. In addition, in order to attribute bath
characteristics to the environment of quantum harmonic oscillators it is
customary to replace the sum over $j$ by an integral over a continuous set
of frequencies $\omega $\ and effect the following scaling:
\begin{equation}
\sum_{j}=N\int_{0}^{\infty }d\omega J(\omega ),~~m_{j}=m,~~C_{j}=\frac{C}{%
\sqrt{N}}\,,
\end{equation}%
where $J(\omega )$ is the so-called `spectral density'. Thus, the
fluctuation in the work is given by
\begin{equation}
\sigma ^{2}=\int_{0}^{\infty }d\omega J(\omega )\frac{C^{2}\hbar }{m\omega }%
\coth \left( \frac{1}{2}\hbar \beta \omega \right) \times [|\beta _{\omega
}^{2}(t)|+|\bar{\beta}_{\omega }^{2}(t)|+|\delta _{\omega }^{2}(t)|+|\bar{%
\delta}_{\omega }^{2}(t)|]\,.
\end{equation}%
This expression is much too complicated to make any headway in our effort to
relate $\langle W^{2}(t)\rangle $ to $\langle W(t)\rangle $, unlike the
classical case. We therefore invoke the Ohmic form of the spectral density
that reads:
\begin{eqnarray}
J(\omega ) &=&\frac{3\omega ^{2}}{\omega _{D}^{3}},~~\omega \leq \omega
_{D}\,,  \notag \\
&=&0,~~\omega \geq \omega _{D}\,,
\end{eqnarray}%
where $\omega _{D}$ is a high-frequency cut-off. Further, the friction $%
\gamma $ is defined as the lumped parameter:
\begin{equation}
\gamma =\frac{3C^{2}\pi }{\omega _{D}^{3}}\,.
\end{equation}%
Hence,
\begin{equation}
\sigma ^{2}=\frac{\gamma }{m\pi }\int_{0}^{\omega _{D}}d\omega \hbar \omega
\coth \left( \frac{\beta \hbar \omega }{2}\right) \times [|\beta _{\omega
}^{2}(t)|+|\bar{\beta}_{\omega }^{2}(t)|+|\delta _{\omega }^{2}(t)|+|\bar{%
\delta}_{\omega }^{2}(t)|]\,,  \label{48}
\end{equation}%
The equation (\ref{48}) allows for assessing the fluctuation in the work
operator for a wide range of parameters that characterise the external
force-field. This will be demonstrated in section IV in terms of an explicit
choice of the force $f(t)$.

We note that for ohmic dissipation cf. Eqs. (\ref{bt}) and (\ref{dt}),
\begin{equation}
B_{\omega }(t)=\int_{0}^{t}d\tau e^{-(\gamma +i\omega _{c})(t-\tau )}\cos
\omega \tau \,,
\end{equation}%
\begin{equation}
D_{\omega }(t)=\int_{0}^{t}d\tau e^{-(\gamma +i\omega _{c})(t-\tau ^{\prime
})}\sin \omega \tau ^{\prime }\,.
\end{equation}%
Therefore,
\begin{equation}
\beta _{\omega }(t)=\int_{0}^{t}dt_{1}f(t_{1})\int_{0}^{t_{1}}d\tau
e^{-(\gamma +i\omega _{c})(t_{1}-\tau )}\cos \omega \tau \,,
\end{equation}%
\begin{equation}
\delta _{\omega }(t)=\int_{0}^{t}dt_{2}f(t_{2})\int_{0}^{t_{2}}d\tau
^{\prime }e^{-(\gamma +i\omega _{c})(t_{2}-\tau ^{\prime })}\sin \omega \tau
^{\prime }\,.
\end{equation}%
It follows then
\begin{eqnarray}
|\beta _{\omega }(t)|^{2} &=&\beta _{\omega }(t)\beta _{\omega }^{\ast
}(t)=\int_{0}^{t}dt_{1}f(t_{1})\int_{0}^{t}dt_{2}f(t_{2})  \notag \\
&\times &\int_{0}^{t_{1}}d\tau e^{-(\gamma +i\omega _{c})(t_{1}-\tau )}\cos
\omega \tau \int_{0}^{t_{2}}d\tau ^{\prime }e^{-(\gamma -i\omega
_{c})(t_{2}-\tau ^{\prime })}\cos \omega \tau ^{\prime }\,.  \label{53}
\end{eqnarray}%
Similarly,
\begin{eqnarray}
|\bar{\beta}_{\omega }(t)|^{2}
&=&\int_{0}^{t}dt_{1}f(t_{1})\int_{0}^{t}dt_{2}f(t_{2})  \notag \\
&&\times \int_{0}^{t_{1}}d\tau e^{-(\gamma -i\omega _{c})(t_{1}-\tau )}\cos
\omega \tau \int_{0}^{t_{2}}d\tau ^{\prime }e^{-(\gamma +i\omega
_{c})(t_{2}-\tau ^{\prime })}\cos \omega \tau ^{\prime }\,.  \label{54}
\end{eqnarray}

The corresponding expressions for $|\delta _{\omega }(t)|^{2}$ and $|\bar{%
\delta}_{\omega }(t)|^{2}$ are obtained by replacing the cosine functions by
sine functions, in Eqs. (\ref{53}) and (\ref{54}) respectively. Combining
all these terms we obtain from Eq. (\ref{48}),
\begin{eqnarray}
\sigma ^{2}&=&\frac{\gamma }{m\pi }\int_{0}^{\omega _{D}}d\omega \hbar
\omega \coth \left( \frac{\beta \hbar \omega }{2}\right) \times
\int_{0}^{t}dt_{1}f(t_{1})\int_{0}^{t}dt_{2}f(t_{2})\int_{0}^{t_{1}}d\tau
\int_{0}^{t_{2}}d\tau ^{\prime }\cos [\omega (\tau -\tau ^{\prime })]  \notag
\\
&&\times e^{\gamma (\tau +\tau ^{\prime })-\gamma (t_{1}+t_{2})}\cos [\omega
_{c}(\tau -\tau ^{\prime }-t_{1}+t_{2})].  \label{55}
\end{eqnarray}%
Even after incorporating the Ohmic dissipation model for the spectral
function, the above expression is much too complicated to arrive at a
closed-form structure. Hence, we employ an often-used weak-coupling/low-friction approximation, familiar in Quantum Optics, that variantly goes
under the name of the `Rotating Wave Approximation' (RWA) \cite{22newgsabook,23newscu}
which implies that the dominant frequency to be picked up from the spectrum
of the bath is the cyclotron frequency $\omega _{c}$. Thus the regime of
validity of this approximation is the one in which $(\gamma /\omega _{c}\ll
1)$, as would happen when the external magnetic field is sufficiently large,
as in the quantum Hall regime. Given this premise we note that the two
cosine functions in Eq. (\ref{55}), when combined and integrated over $\tau $
and $\tau ^{\prime }$ would yield Lorentzians of the form:
\begin{equation}
\frac{\gamma }{\gamma ^{2}+(\omega \pm \omega _{c})^{2}}\,,
\end{equation}%
which, under the RWA, lead to delta functions $\pi \delta (\omega \pm \omega
_{c})$. The net result, following a time ordering of $t_{1}>t_{2}$, amounts
to
\begin{eqnarray}
\langle \sigma ^{2}(t)\rangle &=&\frac{1}{m}\hbar \omega _{c}\coth \left(
\frac{1}{2}\hbar \beta \omega _{c}\right)  \notag \\
&&\int_{0}^{t}dt_{1}f(t_{1})\int_{0}^{t_{1}}dt_{2}f(t_{2})e^{-\gamma
(t_{1}+t_{2})}\cos \omega _{c}(t_{1}-t_{2})\int_{0}^{t_{2}}e^{2\gamma \tau
^{\prime }}d\tau ^{\prime }  \notag \\
&=&\frac{\hbar \omega _{c}}{m}\coth \left( \frac{1}{2}\hbar \beta \omega
_{c}\right) \times
\int_{0}^{t}dt_{1}f(t_{1})\int_{0}^{t_{1}}dt_{2}f(t_{2})e^{-\gamma
(t_{1}-t_{2})}\cos \omega _{c}(t_{1}-t_{2})\,,
\end{eqnarray}%
where the transient term, proportional to $\exp [-\gamma (t_{1}+t_{2})]$ has
been dropped and the cut-off $\omega _{D}$\ has been tacitly taken to be
much larger than the cyclotron frequency $\omega _{c}$. We thus have the
desired result:
\begin{equation}
\langle \sigma ^{2}(t)\rangle =\hbar \omega _{c}\coth \left( \frac{\beta
\hbar \omega _{c}}{2}\right) \langle W(t)\rangle \,.  \label{final}
\end{equation}%
Evidently, in the classical limit the prefactor on the right of Eq. (\ref%
{final}) reduces to $2KT$, as expected.

It is important to point out that not only do we have a relation connecting $%
\langle W^{2}(t)\rangle $ to $\langle W(t)\rangle $ we have, in fact, a
closed-form expression for the characteristic function itself, that reads
\begin{equation}
C_{W}(t)=\exp \left\{ -\frac{h^{2}}{2}\hbar \omega _{c}\coth \left( \frac{%
\beta \hbar \omega_{c} }{2}\right) \langle W(t)\rangle \right\} \times e^{ih
\phi (t)}.
\end{equation}%
Because $\langle W^{n}(t)\rangle =(-i)^{n}\lim_{h\rightarrow 0}\frac{d^{n}}{%
dh^{n}}C_{W}(t)$, all moments of the work operator and hence its probability
distribution are calculable, at least in the RWA.

\section{Further Insights for a specific choice of the external drive}

In this section we consider a special form of $f(t)$ in order to bring out a
few essential quantum features, especially at low temperatures, which make
our analyses stand apart from hitherto treated classical cases in the
context of fluctuation relations. We assume
\begin{equation}
f(t)=f_{0}e^{-\Gamma t}cos\Omega t.  \label{60}
\end{equation}

The choice is dictated by a couple of reasons: (i) for an unbounded force $%
f(t)$ the mean work $\langle W\rangle $ diverges in the long time limit--the
damping parameter $\Gamma $ prevents it from becoming so; (ii) the driving
frequency $\Omega $ provides an additional handle to tune the system vis a
vis the intrinsic cyclotron frequency $\omega _{c}$.

\subsection{The long-time limit}

With this form of $f(t)$ in Eq. (\ref{60}) we first calculate the mean work $%
\langle W(t)\rangle $. Using Eq. (\ref{40}), the properties of exponential
functions and some algebra, we find, in the long-time-limit.

\begin{align}
& \!\!\lim_{t\rightarrow \infty }\langle W(t)\rangle =\frac{f_{0}^{2}}{4m}\{%
\frac{1}{(\gamma -\Gamma )-i(\omega _{c}+\Omega )}\left[ \frac{1}{2}(\frac{1%
}{\Gamma }+\frac{1}{\Gamma -i\Omega })-\frac{1}{\gamma +\Gamma -i(\omega
_{c}+\Omega )}-\frac{1}{\gamma +\Gamma -i(\omega _{c}-\Omega )}\right]
\notag \\
& +\frac{1}{(\gamma -\Gamma )+i(\omega _{c}-\Omega )}\left[ \frac{1}{2}(%
\frac{1}{\Gamma }+\frac{1}{\Gamma -i\Omega })-\frac{1}{\gamma +\Gamma
+i(\omega _{c}-\Omega )}-\frac{1}{\gamma +\Gamma +i(\omega _{c}+\Omega )}%
\right] \}+c.c.\,.  \notag \\
&
\end{align}

It is clear that the driving frequency can be varied, as stated earlier. Two
limiting cases, detailed below, deserve special attention: \newline
Case 1. $\Omega =0$ \newline

We derive
\begin{equation}
\lim_{t\rightarrow \infty }\!\!\langle \hat{W}(t)\rangle =\frac{f_{0}^{2}}{%
m\Gamma }\frac{(\Gamma +\gamma )}{(\Gamma +\gamma )^{2}+\omega _{c}^{2}}.
\label{62}
\end{equation}%
Case 2. $\Omega \rightarrow \omega _{c}$ \newline

As $\Omega $ approaches $\omega _{c}$ the leading contribution in Eq. (\ref%
{62}) yields
\begin{equation}
\lim_{t\rightarrow \infty }\langle W(t)\rangle =\frac{f_{0}^{2}}{4m\Gamma }%
\frac{(\Gamma +\gamma )}{(\Gamma +\gamma )^{2}+(\Omega -\omega _{c})^{2}}.
\end{equation}

Our next task is to compute $\langle W^{2}(t)\rangle $ from Eq. (\ref{48}).
Towards this end, we need to calculate the term within the square
parentheses in Eq. (\ref{48}). Evidently,
\begin{equation}
\lbrack \left\vert \beta _{\omega }^{2}(t)\right\vert +\left\vert \delta
_{\omega }^{2}(t)\right\vert +\left\vert \bar{\beta _{\omega }^{2}}%
(t)\right\vert +\left\vert \bar{\delta _{\omega }^{2}}(t)\right\vert
]=[\left\vert \beta _{\omega }-i\delta _{\omega }\right\vert ^{2}+\omega
_{c}\leftrightarrow -\omega _{c}].
\end{equation}

Now, from Eqs. (\ref{A}) and (\ref{bt}) and the special form of $A(z)$ in
Eq. (\ref{13}) in the Ohmic limit
\begin{equation}
\beta _{\omega }-i\delta _{\omega
}=\int_{0}^{t}dt_{1}f(t_{1})\int_{0}^{t_{1}}d\tau e^{-(\gamma +i\omega
_{c})(t_{1}-\tau )}e^{-i\omega \tau }.  \label{65}
\end{equation}

After a few steps we obtain%
\begin{equation}
\beta _{\omega }-i\delta _{\omega }=\frac{f_{0}}{2\left[ \gamma +i(\omega
_{c}-\omega )\right] }\{\frac{1-e^{-t[\Gamma -i(\Omega -\omega )]}}{\Gamma
-i(\Omega -\omega )}-\frac{1-e^{-t[\Gamma +\gamma -i(\Omega -\omega _{c})]}}{%
\Gamma +\gamma -i(\Omega -\omega _{c})}\}+\Omega \leftrightarrow -\Omega .
\end{equation}

Thus in the $t\rightarrow \infty $ limit,
\begin{equation}
\left\vert \beta _{\omega }-i\delta _{\omega }\right\vert ^{2}=\frac{%
f_{0}^{2}}{4}\left\vert \bigg[\frac{1}{[\Gamma -i(\Omega -\omega )][(\Gamma
+\gamma )-i(\Omega -\omega _{c})]}+\frac{1}{[\Gamma +i(\Omega +\omega
)][(\Gamma +\gamma )+i(\Omega +\omega _{c})]}\bigg]\right\vert ^{2}.
\end{equation}

We now deal with the two limiting cases $\Omega =0$ and $\Omega =\omega _{c}$
separately.\newline
Case 1: $\Omega =0$
\begin{equation}
\left\vert \beta _{\omega }-i\delta _{\omega }\right\vert ^{2}+\left\vert
\bar{\beta _{\omega }}-i\bar{\delta _{\omega }}\right\vert ^{2}=\frac{%
2m\Gamma }{(\Gamma +\gamma )}.\frac{1}{\Gamma ^{2}+\omega ^{2}}\langle
W\rangle .
\end{equation}%
Substituting in Eq. (\ref{48}), we find for $\Omega =0$%
\begin{equation}
\langle \sigma ^{2}\rangle =\frac{2 \gamma \Gamma }{\pi (\Gamma +\gamma )}%
\left\langle W\right\rangle \int_{0}^{\infty }d \omega \frac{\hbar \omega }{%
(\Gamma ^{2}+\omega ^{2})}coth\left( \frac{1}{2}\hbar \beta \omega \right).
\end{equation}
Case 2: Next we find, as $\Omega $ approaches $\omega _{c}$, up to leading
orders,
\begin{equation}
\left\vert \beta _{\omega }-i\delta _{\omega }\right\vert ^{2}+\left\vert
\bar{\beta} _{\omega }-i\bar{\delta} _{\omega }\right\vert ^{2}=\frac{%
m\Gamma }{(\Gamma +\gamma )}\frac{1}{\Gamma ^{2}+(\omega _{c}-\omega )^{2}}%
\langle W\rangle.
\end{equation}
Hence,
\begin{equation}
\langle \sigma ^{2}\rangle =\frac{\gamma \Gamma }{\pi (\Gamma +\gamma )}%
\langle W\rangle \int_{0}^{\infty }d\omega \frac{\hbar \omega }{\Gamma
^{2}+(\omega _{c}-\omega )^{2}}coth\left( \frac{1}{2}\hbar \beta \omega
\right).
\end{equation}
In either case we may use the following form of the Dirac delta function:
\begin{equation}
\lim_{\Gamma \rightarrow 0}\frac{\Gamma }{\Gamma ^{2}+(\omega -\nu )^{2}}%
=\pi \delta (\omega -\nu ),
\end{equation}%
to yield
\begin{equation}
\langle \sigma ^{2}\rangle =2\langle W\rangle KT,\text{ for }\Omega =0\text{,%
}  \label{73}
\end{equation}
\begin{equation}
\langle \sigma ^{2}\rangle =\langle W\rangle \hbar \omega _{c}coth\left(
\beta \frac{\hbar \omega _{c}}{2}\right) \text{, for }\Omega =\omega _{c}%
\text{.}  \label{74}
\end{equation}%
The parameter $\alpha$ is then equal to $tanh\left(\beta \frac{\hbar \omega _{c}}{2}\right) /(\frac{\hbar \omega _{c}}{2})$
and hence $\alpha K_B T =tanh\left(\beta \frac{\hbar \omega _{c}}{2}\right) /(\frac{\beta \hbar \omega _{c}}{2}) $. It is
interesting that it depends on the cyclotron freq and which fixes us how large it can be for a given temperature.
Thus, in the high temperature $T\rightarrow \infty $ classical limit, the
second case of Eq. (\ref{74}) merges with the $\Omega =0$ case of Eq. (\ref%
{73}). However, for $\Omega \neq 0$, and in particular, for $\Omega =\omega
_{c}$, the results are quite different. Specifically, for $T = 0$, the case
of $\Omega =\omega _{c}$ leads to
\begin{equation}
\langle \sigma ^{2}\rangle =\hbar \omega _{c}\langle W\rangle ,
\end{equation}
which implies that in the strictly quantum regime, the zero point energy $%
\hbar \omega _{c}$ replaces the thermal energy $(2KT)$ of the classical case
of I.

\subsection{Quantum Fluctuation Theorems in Short Time Limit}

We saw earlier that though in the Ohmic dissipation limit the quantum
Langevin equation can be described in terms of a constant friction the noise
is still `colored' i.e. non-white rendering non-Markovianess into the
system. It is well known that these non-Markovian effects are most prominent
when the system dynamics is over a timescale smaller than a typical
`quantum' time defined by $\hbar /KT$. Thus at low temperatures this
characteristic quantal time scale can be appreciably large, within which
quantum phenomena would be most effective. In order to capture these
strictly quantum effects it is necessary to consider the short time limits
of $\left\langle W\right\rangle $\ and $\sigma ^{2}$. Concomitantly, it is
also important to relax the condition on the bath cut-off frequency $\omega
_{D}$ and take it to be finite. With this preamble we now take up the task
of evaluating $\left\langle W\right\rangle $\ and $\sigma ^{2}$ in the short
time limit \cite{31tra}. As far as $\left\langle W\right\rangle $ is
concerned we can either look at the explicit algebraical expression of Eq. (%
\ref{40}) emanating from the stated structure of the driving force as is Eq.
(\ref{60}) or surmise directly from the double integral form of Eq. (\ref{40}%
) that to leading order, in the short-time limit,%
\begin{equation}
\left\langle W\right\rangle =\frac{f_{0}^{2}}{2m}t^{2}.
\end{equation}
Next, we find from Eq. (\ref{65}) that%
\begin{equation}
\lim_{t\rightarrow 0}\left( \beta _{\omega }-i\delta _{\omega }\right) =%
\frac{f_{0}}{2}t^{2}.
\end{equation}
Hence,%
\begin{equation}
\left\vert \beta _{\omega }-i\delta _{\omega }\right\vert ^{2}+\left\vert
\overline{\beta }_{\omega }-i\overline{\delta }_{\omega }\right\vert ^{2}=%
\frac{f_{0}^{2}}{2}t^{4}.
\end{equation}
Substituting in Eq. (\ref{48})%
\begin{equation}
\sigma ^{2}=\frac{\gamma f_{0}^{2}t^{4}}{8m\pi }\int_{0}^{\omega _{D}}\hbar
\omega \coth \left( \frac{\beta \hbar \omega }{2}\right) ,
\end{equation}
which, in the further low-temperature limit, reduces to%
\begin{equation}
\lim_{T\rightarrow 0}\sigma ^{2}=\frac{\hbar \gamma }{16m\pi }%
f_{0}^{2}t^{4}\omega _{D}.
\end{equation}
Therefore,
\begin{equation}
\lim_{T\rightarrow 0}\sigma ^{2}=\left\langle W^{2}\right\rangle =\left(
\frac{\hbar \gamma }{8m\pi }\right) \left( \omega _{D}t\right)
^{2}\left\langle W\right\rangle ,
\end{equation}%
thus providing a relation between $\left\langle W^{2}\right\rangle $ and $%
\left\langle W\right\rangle $ in the low-temperature, short-time limit.

\section{Concluding remarks}

The subject of fluctuation theorems has a fairly long history, yet there has
been a flurry of activities in the last few years, mainly in the classical
biological systems. The foundational issues of quantum fluctuation relations
have also attracted attention in the last decade or so.

The most widely studied problem of quantum diffusion that is amenable to
explicit calculation is the Brownian motion of a one dimensional quantum
harmonic oscillator \cite{29sdgdiffusion}. The present problem of
dissipative cyclotron motion is however elevated to a higher dimensional
space of four, as it involves two coordinates and two momenta in a plane
normal to the direction of the applied magnetic field. Thus, it is of
potential interest in the quantum version of what has been dubbed as
isometric fluctuations, when diffusive motions in different dimensions are
asymmetric \cite{32kum}.

The above study, though couched in the contextual domain of orbital
magnetism, is likely to have widespread applications to a large class of
problems in Quantum Optics. We hope to make further investigations of
quantum fluctuation relations for two and higher level systems which fall
under the generic nomenclature of spin-boson Hamiltonians \cite{29sdgdiffusion}.

\section*{Acknowledgements}

GSA thanks the Tata Institute of Fundamental Research, Mumbai where a part
of this work was done. SD is grateful to the Department of Science and
Technology for its support through the J.C. Bose Fellowship. We thank Dr.
Jishad Kumar and Mr. Chiranjib Barman for their help in preparation of the
manuscript.

\end{document}